\documentclass{vienna-conf2019}
\usepackage{graphicx}
\usepackage{hyperref}
\usepackage[]{natbib}  
\usepackage{epstopdf}

\def\BibTeX{{\rm B\kern-.05em{\sc i\kern-.025em b}\kern-.08em
    T\kern-.1667em\lower.7ex\hbox{E}\kern-.125emX}}
\bibpunct{(}{)}{;}{a}{}{,}  

\begin{document}

\TitreGlobal{Stars and their variability observed from space}


\title{Listening to the Heartbeat: Tidal Asteroseismology in Action}

\runningtitle{Tidal Asteroseismology}

\author{Z. Guo}\address{Department of Astronomy and Astrophysics, 525 Davey Lab, The Pennsylvania State University, USA}





\setcounter{page}{237}


\maketitle


\begin{abstract}
We briefly review the current status of the study of tidally excited oscillations (TEOs) in heartbeat binary stars. Particular attention is paid to correctly extracting the TEOs when the Fourier spectrum also contains other types of pulsations and variabilities. We then focus on the theoretical modeling of the TEO amplitudes and phases. Pulsation amplitude can be modeled by a statistical approach, and pulsation phases can help to identify the azimuthal number $m$ of pulsation modes. We verify the results by an ensemble study of ten systems. We discuss some future prospects, including the secular evolution and the non-linear effect of TEOs.  
\end{abstract}



\section{Introduction}


More than half of all stars reside in binaries, and tides can 
have a significant effect on stellar oscillations. In the first version of the classical textbook  {\it Nonradial oscillations of stars}, there is a whole chapter on tidal oscillations \citep{Unno79}. However, it was removed in the second version \citep{Unno89}, probably owing to the notion that such oscillations are difficult to be observed in practice. 

The direct manifestation of equilibrium and dynamical tides can be shown in the light curves as flux variations. The theoretical foundations were laid out several decades ago, and \citet{kum95} even derived an expression for the observed flux variations from the tidal response. However, it is only after the {\it Kepler} satellite that we are able to observe unambiguously the tidally excited oscillations. The prototype system KOI-54 (Welsh et al. 2011), inspired lots of interests in observational (\citealt{ham13,ham16,ham18}; \citealp{guo17,guo19a}) and theoretical studies (\citealp{Fuller12};  \citealp{bur12}; \citealp{Fuller13}; \citealp{ole14}; \citealp{Fuller17a};
\citealp{pen19}). \citet{tho12} presented tens of such so-called `heartbeat' stars (HBs) and the Kepler Eclipsing Binary (EB) Catalog (\citealt{kir16}) now consists of 173 such systems, flagged as `HBs'. Out of them, about 24 systems show tidally excited oscillations (TEOs), flagged as `TPs'. Spectroscopic follow-up observations are accumulating (\citealp{smu15}; \citealp{shp16}). Other space missions such as BRITE revealed more massive HBs with TEOs (\citealt{pig18}), including the O-type binary $\iota$ Ori (\citealt{pab17}). The first sector data of the ongoing TESS mission already offered us a massive HB with TEOs (\citealt{jay19}).

\section{Extracting Tidally Excited Oscillations from Binary Light Curves}

To study the TEOs, we usually first model the flux variations from the equilibrium tide. Dedicated light curve synthesis codes are usually used such as the Wilson-Devinney \citep{wil71} and its further development in PHOEBE \citep{prs05}.
In Figure 1, we show the binary light curve models as red lines. For low inclination systems, the light curves usually contain only one periastron brightening bump. If the binary orbit has a moderate inclination, both a hump and a dip are present. For high inclination systems, the light curves usually contain both a bump and an eclipse. The binary models presented here are either from the simplified light curve model (Kumar model for KIC 9016693 and KIC 5034333) or from dedicated synthesis codes (KIC 4142768).

After subtracting the binary model, Fourier spectra of the residuals are presented in the right half of Fig 1. Both the low-frequency (g-mode) and high-frequency (p-mode) region can show variabilities.
\begin{figure}[ht!]
 \centering
 \includegraphics[width=0.95\textwidth,clip]{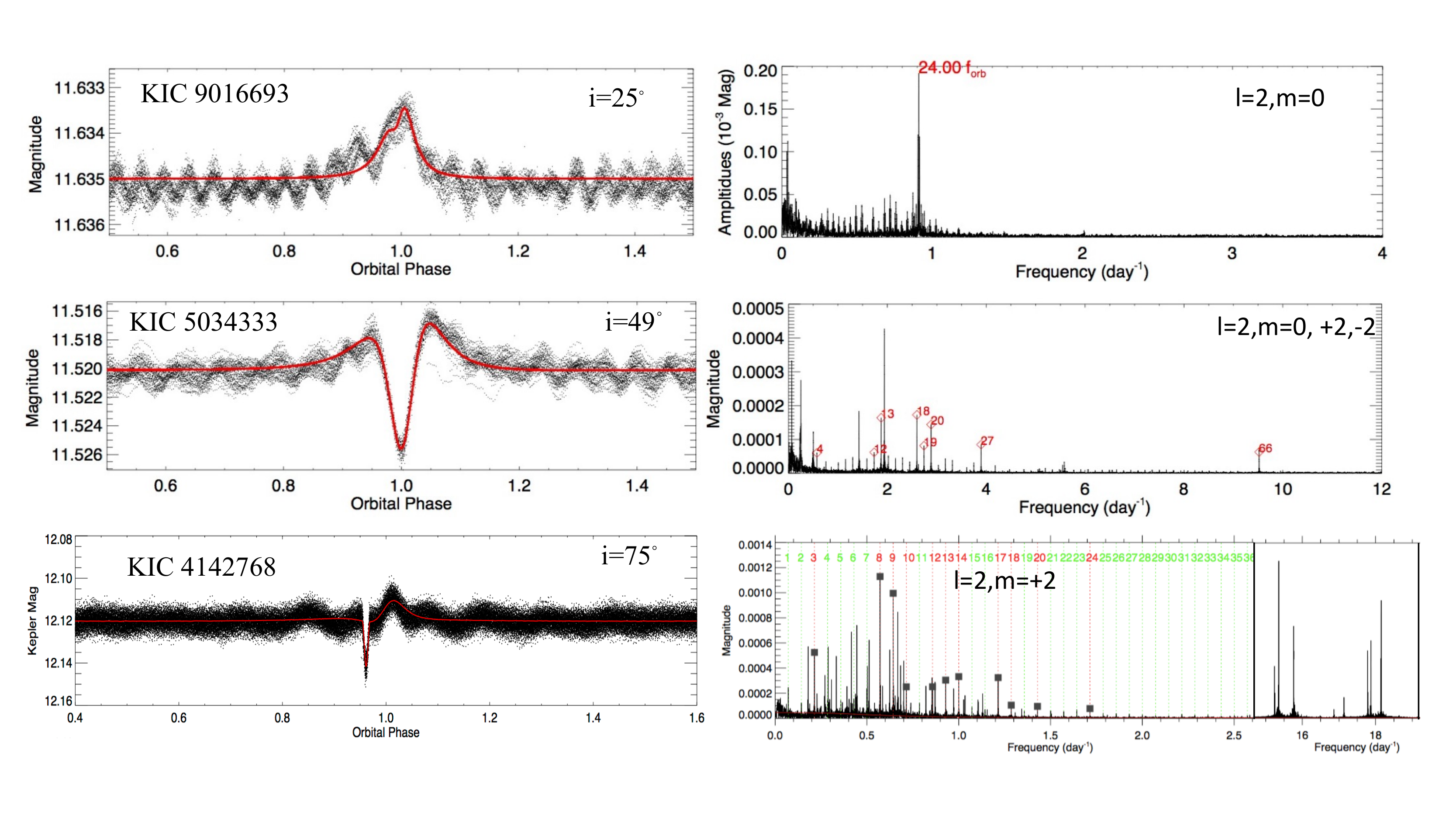}      
  \caption{Phase-folded Kepler light curves of heartbeat binaries and their Fourier spectra. Here we show three typical systems of low, moderate, and high orbital inclinations: from top to bottom, $i =25.6^{\circ}, 49.9^{\circ},$ and $75.8^{\circ}$. The tidally excited oscillations and their mode identifications ($l$ and $m$) are labeled in the Fourier spectra.}
  \label{author1:fig3}
\end{figure}

\subsection{Low-frequency Region}
The tidal forcing frequencies from the companion star naturally fall within the low frequency regime. We thus expect the prominent TEOs are low-frequency g modes\footnote{However, refer to Fuller et al. (2013) for tidally excited p modes in a triple system. In principle, rossby modes can also be tidally excited.} with $l=2$, which are almost always orbital harmonics\footnote{Non-linear effect can generate aharmonics TEOs, e.g., those in KOI-54 and KIC 3230227.}. However, in the low-frequency region of the Fourier spectrum, the variabilities can also arise from the imperfect binary light curve removal (a series of orbital harmonic frequencies), rotational modulations (usually one or two times of the orbital frequency), and $\gamma$ Dor-type self-excited g modes (usually not orbital harmonics, but quasi-linearly spaced in pulsation period). Thus the analysis has to be performed with care.
In the right panels of Fig 1, we show three examples. The TEOs have been labeled, and other variablities are mostly from the rotational signal and the $\gamma$ Dor type g modes.



\subsection{High-frequency Region}
Tidally excited oscillations can coexist with high-frequency p modes. These p-modes can be affected by tides. For instance, the $\delta$ Scuti type p modes in KIC 4544587 are coupled to the tidally excited g-modes, showing a regular pattern spaced by the orbital frequency \citep{ham13}. In the circular, close binary HD 74423, the p modes from the near-Roche-lobe filling primary star are magnified and trapped around the inner L1 point (Handler et al., submitted). The pulsation axis is aligned with the tidal force so that the observed p modes show amplitude and phase variations. In the Fourier spectrum, the perturbed p modes also appear as splittings with a spacing of the orbital frequency.




\section{Modeling the amplitudes and phases of TEOs}

\subsection{Pulsation Amplitude: the Statistical Approach by Fuller (2017)}
Unlike self-excited oscillations, linear theory can predict the amplitude of the tidally forced oscillations. This can be achieved by directly solving the forced oscillation equations (\citealp{bur12}; \citealp{val13}) or by using the mode decomposition method (\citealp{sch02}; \citealp{Fuller12}). The latter method has been used to calculate the amplitude and phases of TEOs with rotation, nonadiabaticity, and spin-orbit misalignment taken into account (\citealp{Fuller17a}).
The pulsation amplitude sensitively depends on the detuning parameter, i.e., the different between the forcing frequency and the eigen-frequency of the star, which cannot be determined accurately. It is better to model the TEO amplitude by a statistical approach, treating the detuning as a random variable. Fig 2 shows the TEO amplitude modeling for KIC 4142768 \citep{guo19a}. The shaded region indicates the $\pm2 \sigma$ credible region, matching the observed amplitudes (gray squares) very well.

For the modeling of TEOs in resonance locking, special treatment is required. Please refer to section 5.1 of \citet{Fuller17a}. Practical applications can be found in \cite{Fuller17b} for KIC 8164262, and Cheng (2020, in prep.) for KIC 11494130.
\begin{figure}[ht!]
 \centering
 \includegraphics[width=0.95\textwidth,clip]{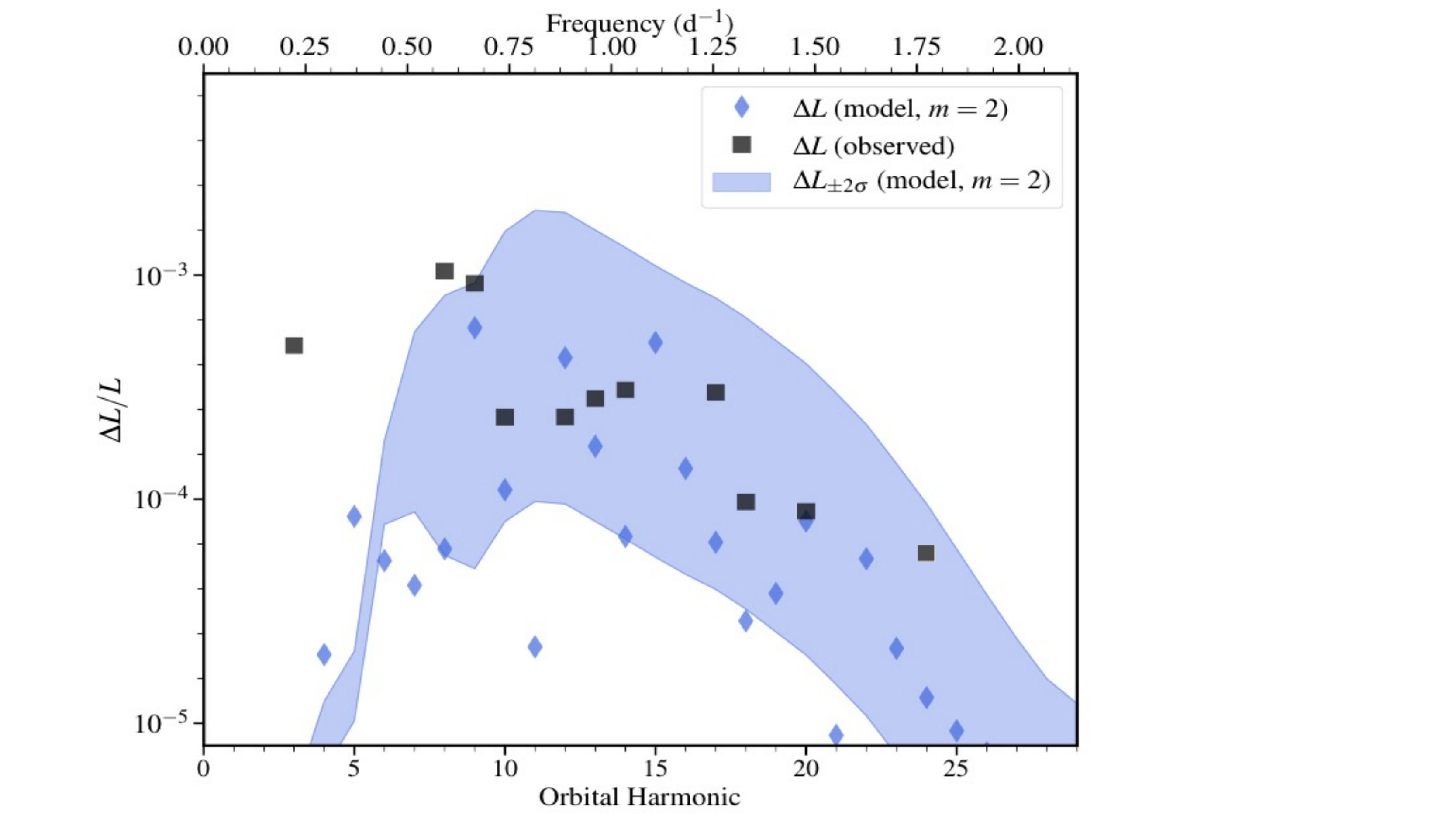}      
  \caption{TEO amplitude modeling of KIC 4142768 assuming a mode identification of $l=2,m=2$. See \citet{guo19a} for details.}
  \label{author1:fig3}
\end{figure}

\subsection{Mode Identification from Pulsation Phases}
Simply speaking, the theoretical adiabatic pulsation phase of TEOs only depends only on the argument of periastron passage $\omega_p$ by the relation $\phi_{l=2, m} =
0.25+m[0.25-\omega_p/(2\pi)] $.
Fig 3 shows the observed TEO phases of six heartbeat binaries (symbols) and the theoretical phases (vertical strips) with the corresponding azimuthal number $m$ labeled. The diagrams of KOI-54 and KIC 3230227 are taken from \citet{ole14}, and \citet{guo17}, respectively. It is encouraging to see that they almost all agree within one sigma. 

However, as shown in Fig. 4, we also find a large number of heartbeat systems that show significant deviations from the theoretical adiabatic phases (see \citealp{guo19b} for more details). The mode non-adiabacticiy and spin-orbit misalignment are likely the reasons behind. \citet{guo19a} considered the mode nonadiabaticity in the theoretical TEO phase calculations, and the theory seems to agree with observations. 

We find that low-inclination systems favor the presence of $m=0$ modes, and very high-inclination systems are more likely to show $|m|=2$ modes. This is in agreement with the simple geometric interpretation.

\begin{figure}[ht!]
 \centering
 \includegraphics[width=0.8\textwidth,clip]{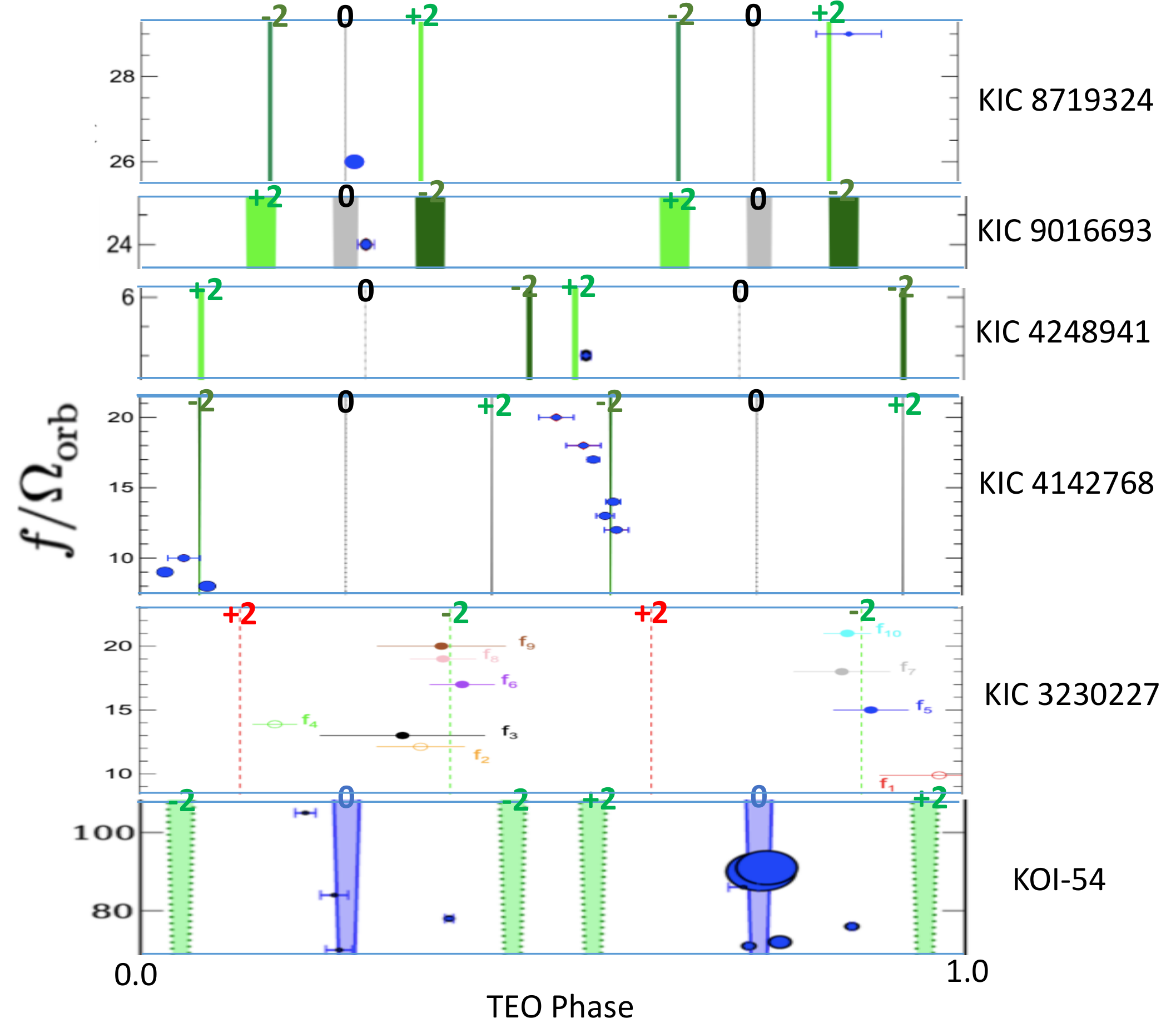}      
 \caption{Pulsation frequencies of TEOs in units of the orbital frequency ($N=f/\Omega_{\rm orb}$) as a function of TEO phases (in units of $2\pi$). These are six heartbeat binaries showing good match with theoretical predictions for TEO phases. The azimuthal number $m$ is labeled.}
 \label{author1:fig1}
\end{figure}

\begin{figure}[ht!]
 \centering
 \includegraphics[width=0.8\textwidth,clip]{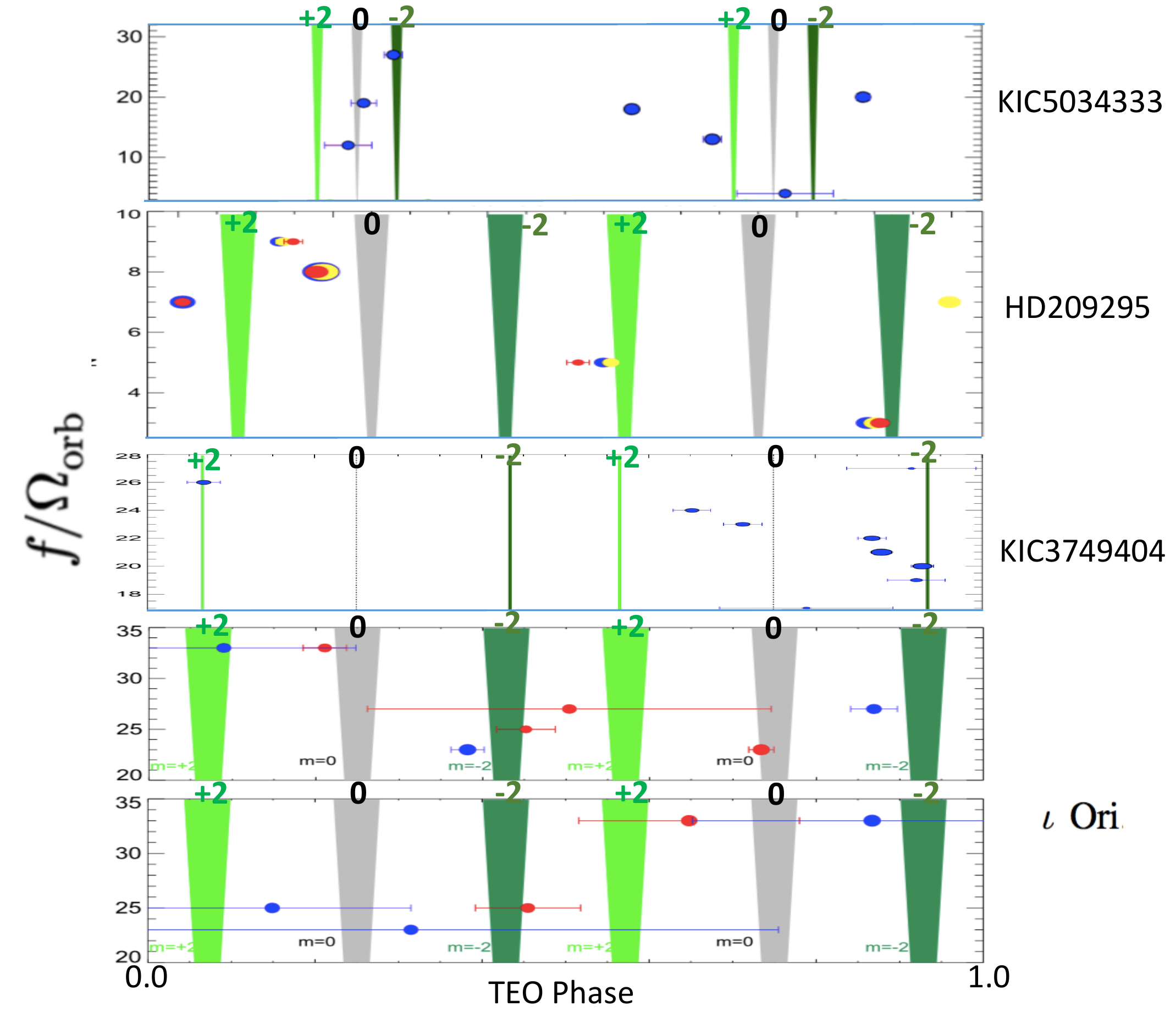}      
  \caption{Same as Fig. 3, but for four systems with significant deviations from the theoretical adiabatic phases (vertical strips).}
  \label{author1:fig2}
\end{figure}


\section{Discussions and Future Prospects}
Other than the mode identification, variations of the amplitude and phase of TEOs can offer us information on the mode damping and the orbital evolution. For KOI-54, it was found that the TEO amplitudes decrease by about $2-3\%$ over three years, which cannot be explained solely by the radiative mode damping \citep{ole14}. A careful observation of these TEOs can identify modes that are undergoing three or multi-mode coupling. Recently, Guo (2019, submitted) found that the non-linear mode coupling in KIC\ 3230227 has probably settled to the equilibrium state. By utilizing the amplitude equations, a detailed analysis of the non-linear mode coupling can be performed \citep{wei12} and is highly desirable. Besides the tidal effect, heartbeat stars are also laboratories for physical processes such as the high-eccentricity orbital migration and precession. Their potential has not been fully exploited.

\begin{acknowledgements}
I thank the organizers for the invitation, and I am also grateful to Jim Fuller for presenting this work on my behalf. We thank Phil Arras and Nevin Weinberg for helpful discussions on nonlinear tides. We thank Rich Townsend for the development of the GYRE oscillation code.
\end{acknowledgements}

\bibliographystyle{aa}  
\bibliography{Guo_1i03} 

\end{document}